\documentclass[sigconf,anonymous=False,review=False]{acmart}
\usepackage{multirow}
\usepackage{makecell}
\usepackage{subfig}
\usepackage{bbding}
\usepackage{enumitem}
\usepackage{colortbl}
\usepackage{diagbox}
\usepackage{comment}
\usepackage{amsmath}
\usepackage{algorithm}
\usepackage{algpseudocode}

\AtBeginDocument{%
  }
\usepackage{newfloat}
\usepackage{listings}

\acmISBN{978-1-4503-XXXX-X/2018/06}

\begin{document}

\title{HeMix: Scaling Industrial Ranking Models with Heterogeneous Token Mixing}

\author{Fangye Wang}
\authornote{Contributed equally to this research.}
\affiliation{%
  \institution{AMAP, Alibaba Group}
  \city{Beijing}
  \country{China}
}
\email{wangfangye.wfy@alibaba-inc.com}

\author{Guowei Yang}
\authornotemark[1]
\affiliation{%
  \institution{AMAP, Alibaba Group}
  \city{Beijing}
  \country{China}
}
\email{liuyi.ygw@alibaba-inc.com}

\author{Xiaojiang Zhou}
\authornote{Corresponding Author}
\affiliation{%
  \institution{AMAP, Alibaba Group}
  \city{Beijing}
  \country{China}
}
\email{zhouxiaojiang.zxj@taobao.com}

\author{Song Yang}
\affiliation{%
  \institution{AMAP, Alibaba Group}
  \city{Beijing}
  \country{China}
}
\email{song.yangs@alibaba-inc.com}

\author{Pengjie Wang}
\affiliation{%
  \institution{AMAP, Alibaba Group}
  \city{Beijing}
  \country{China}
}
\email{pengjie.wpj@alibaba-inc.com}

\begin{abstract}
Scaling up ranking models for industrial recommender systems faces two critical challenges: (C1) existing sequence tokenization fails to jointly capture context-aware and context-invariant user intent from heterogeneous behavior sources, and (C2) prevailing interaction mechanisms are both computationally expensive and semantically homogeneous, limiting prediction quality under strict online latency constraints. We propose \textbf{HeMix}, a scalable ranking model that unifies query-mixed sequence tokenization with heterogeneous feature interaction. To address (C1), HeMix introduces a \textit{Query-Mixed Interest Extraction} module that employs dynamic and fixed queries to simultaneously model context-aware and context-invariant interests from global and real-time behavior sequences. To address (C2), we design the \textit{HeteroMixer} block, comprising Multi-Head Token Fusion, Heterogeneous Mixed-Token Interaction and Group-Aligned Reconstruction, as an efficient alternative to self-attention that enables multi-granularity cross-feature modeling at linear cost. Crucially, HeMix scales smoothly from ${\sim}100$M to ${\sim}1500$M parameters by independently expanding block depth and token dimension, yielding steady accuracy gains without architectural redesign. Experiments on industrial-scale data show that HeMix achieves $+1.64\%$ relative CTR-AUC over the DLRM baseline at the ${\sim}100$M scale while requiring fewer GFLOPs than the strongest competitor. Deployed on the AMAP APP, HeMix yields +0.88\% GMV, +2.74\% PV\_CTR and +0.84\% UV\_CVR over the production baseline in online A/B tests. 
\end{abstract}

\ccsdesc[500]{Information systems~Information retrieval}
\ccsdesc[500]{Information systems~Recommender systems}

\keywords{Heterogeneous Interaction, Industrial Recommenders, Scaling Laws}
\maketitle

\section{Introduction}
Recommender systems (RS) are critical infrastructure for large-scale information distribution platforms. State-of-the-art ranking models are predominantly built upon deep ranking architectures, which capture high-order feature interactions by jointly modeling sequential and non-sequential features~\cite{gui2023hiformer,han2025mtgr}. Representative designs include cross networks~\cite{wang2020dcn,wang2017deep}, attention-based models~\cite{wang2022enhancing,xiao2017attentional}, and Transformer-style approaches~\cite{huang2026hyformer,zhai2024actions,gui2023hiformer}. Inspired by the scaling laws observed in large language models~\cite{kaplan2020scaling,achiam2023gpt}, recent efforts have explored scaling recommender models for further performance gains: Wukong~\cite{zhang2024wukong} stacks FM blocks with linear compression; RankMixer~\cite{zhu2025rankmixer} employs hardware-efficient token mixing with token-specific FFNs; and OneTrans~\cite{zhang2025onetrans} unifies sequential and non-sequential features in a single Transformer backbone. However, despite improved accuracy, these methods still face two key challenges that hinder practical deployment at scale.

\textbf{(C1) Inadequate sequence tokenization for multi-source user behaviors.} Existing approaches either compress behavior sequences into a few tokens by target attention, sacrificing fine-grained semantics, or flatten all behaviors into long token sequences (e.g., OneTrans~\cite{zhang2025onetrans}, MTGR~\cite{han2025mtgr}) at prohibitive computational cost. Moreover, most methods extract only context-aware interests, overlooking the structural properties of user sequences, notably the semantic and timeliness gap between long-term and real-time behaviors, which demand distinct modeling.


\textbf{(C2) Homogeneous and inefficient feature interaction design.} Widening the embeddings or MLPs of conventional ranking models brings quickly diminishing returns, since no token-level interaction is added. Existing interaction operators, in turn, are limited in two ways. \textit{First}, they are costly: pairwise attention scales quadratically with the number of tokens, and cross networks need many stacked layers to reach high-order crosses, both straining the online latency budget. \textit{Second}, they are homogeneous: one set of projections is shared by all tokens, and the group-specific projections of recent work~\cite{gui2023hiformer,zhang2025onetrans} still sit inside quadratic attention. Heterogeneity and linear cost have thus not been achieved together.

To address these challenges, we propose \textbf{HeMix}, a scalable ranking model that explicitly captures heterogeneous mixing characteristics in both sequence tokenization and token interaction. For \textbf{C1}, HeMix introduces a \textbf{Query-Mixed Interest Extraction} tokenizer that leverages non-sequential features as dynamic queries, augmented with learnable fixed interest vectors. These queries jointly attend to user behavior sequences through a novel \textbf{Mixed Hetero Attention} mechanism, enabling simultaneous modeling of context-aware and context-invariant interests across both long-term and real-time sequences. For \textbf{C2}, HeMix employs the \textbf{HeteroMixer Block}, comprising Multi-Head Token Fusion, Heterogeneous Mixed-Token Interaction, and Group-Aligned Reconstruction, which replaces quadratic pairwise attention with a \textbf{fuse--interact--reconstruct} pipeline and thereby captures multi-granularity, cross-subspace dependencies among heterogeneous features at linear cost. The architecture supports flexible scaling by varying block depth or token dimensions without redesigning core components. HeMix has been fully deployed on AMAP APP, a leading navigation and location-based service (LBS) platform serving hundreds of millions of daily active users, bringing substantial improvements in core business metrics.


Our main contributions are summarized as follows:
\begin{itemize}[leftmargin=*]
\item We propose HeMix, a scalable ranking model that unifies behavior sequence tokenization and heterogeneous feature interaction in a single architecture.
\item The Query-Mixed Interest Extraction module jointly captures context-aware and context-invariant user interests from both long-term and real-time behavior sequences.
\item The HeteroMixer block, as a linear-complexity alternative to self-attention, enables multi-granularity heterogeneous interaction under online latency constraints.
\item Extensive offline experiments validate the accuracy and scaling behavior of HeMix, and full deployment on the AMAP APP delivers $+0.88\%$ GMV, $+2.74\%$ PV\_CTR and $+0.84\%$ UV\_CVR.
\end{itemize}

\section{Related Work}

\textbf{Dense Interaction Architectures.} Modern recommendation systems are predominantly built upon Deep Learning Recommendation Models (DLRMs), for which effective modeling of feature interactions is a key determinant of performance. Early approaches such as WDL~\cite{cheng2016wide} combine logistic regression and deep neural networks (deep component) to capture low- and high-order interactions, respectively. Subsequent works enhance interaction modeling by integrating structured inductive biases: DeepFM~\cite{guo2017deepfm} fuses Factorization Machines (FM)~\cite{rendle2012factorization}, CIN~\cite{lian2018xdeepfm}, AFN~\cite{cheng2020adaptive} with DNNs; DeepCross leverages residual networks to implicitly learn cross features; while PNN~\cite{qu2018product}, DCNv2~\cite{wang2020dcn}, FinalMLP~\cite{mao2023finalmlp}, xDeepFM~\cite{lian2018xdeepfm}, AFN+~\cite{cheng2020adaptive}, FINAL~\cite{zhu2023final} explicitly design dedicated operators to model high-order interactions. More recently, attention-based architectures like AutoInt~\cite{song2019autoint} and HiFormer~\cite{gui2023hiformer} employ self-attention~\cite{vaswani2017attention} with residual~\cite{he2016resnet} connections to capture complex, adaptive interactions, and DIN~\cite{zhou2018din}, DIEN~\cite{zhou2019dien} further combine the sequence information. Despite higher accuracy, these interaction-heavy models suffer from increased latency and memory use when scaled~\cite{zhu2025rankmixer, wang2023towards}. Worse, blind scaling, e.g., deeper nets or larger embeddings—often yields diminishing returns or even hurts performance, revealing the pitfalls of over-parameterization. This calls for structured, efficient interaction designs over brute-force expansion.

\textbf{Scaling up Recommendation Models.} 
Scaling laws—empirical power-law relationships between model performance and scaling dimensions such as model size, dataset scale, and compute budget—have emerged as a foundational principle in deep learning~\cite{kaplan2020scaling, henighan2020scaling}, driving transformative advances in Natural Language Processing (NLP)~\cite{brown2020language, hoffmann2022training}, Computer Vision (CV)~\cite{touvron2021training, radford2021learning}, and multimodal modeling~\cite{alayrac2022flamingo, radford2021clip} over the past decade. Recently, their applicability to recommender systems has attracted growing interest. Existing efforts explore scaling strategies along several axes: pretraining on large-scale user activity sequences, learning general-purpose user representations, and scaling online retrieval architectures. For instance, Wukong~\cite{zhang2024wukong} stacks factorization machines (FMs) with linear compress blocks(LCBs) to enhance feature interaction modeling. In parallel, Zhang et al.~\cite{zhang2024scaling} scaled a sequential recommendation model to 0.8 billion parameters, demonstrating feasibility at extreme scales. MTGR combines the advantages of DLRM and GRM and introduces group-layer normalization and dynamic masking strategies to achieve better performance. RankMixer~\cite{zhu2025rankmixer} designs multihead token mixing and per-token FFN strategies to capture heterogeneous interactions, and adopts a dynamic routing strategy to improve scalability. This work addresses a key challenge: how to design a recommender architecture that efficiently captures both global and local interactions among semantic tokens while enabling scalable, structured performance gains.

\section{Methodology: HeMix}

\begin{figure*}[t]
\centering
\subfloat[Overview of HeMix.]{
    \begin{minipage}[t]{0.38\linewidth}
    \centering
    \includegraphics[width=0.85\textwidth]{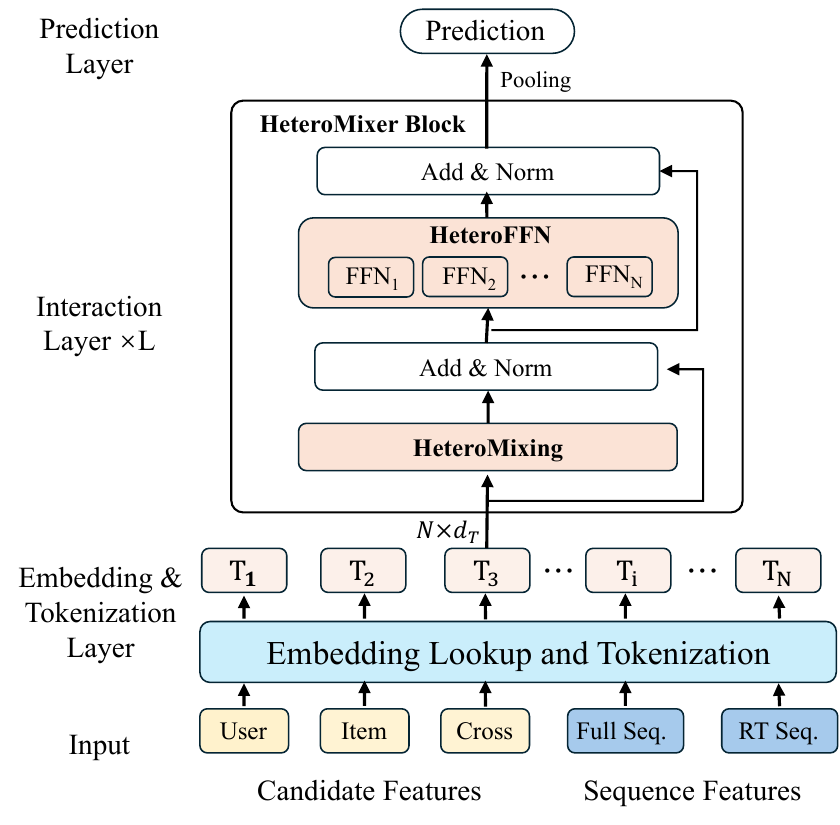}
    \end{minipage}
}
\centering
\subfloat[Feature embedding and tokenization.]{
    \begin{minipage}[t]{0.31\linewidth}
    \centering
    \includegraphics[width=0.85\textwidth]{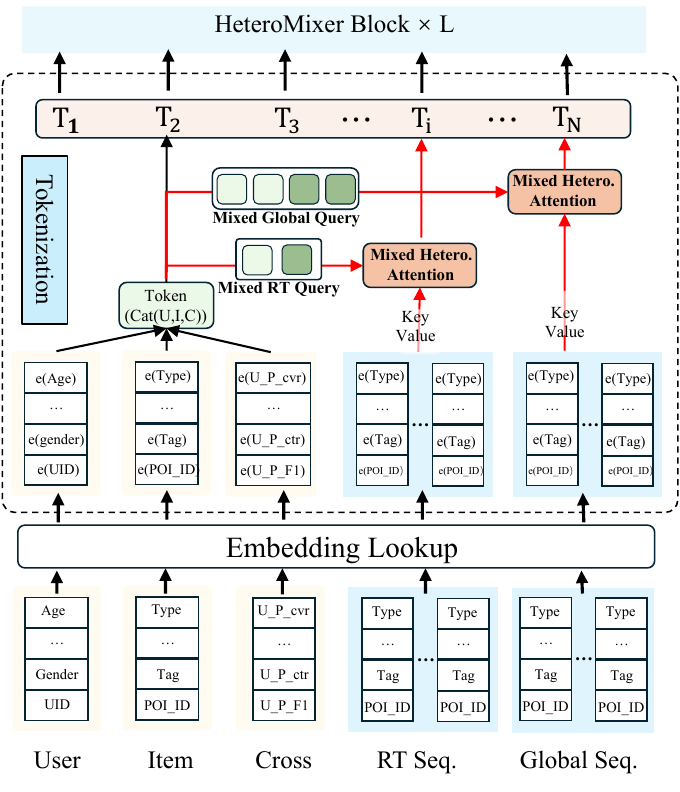}
    \end{minipage}
}
\subfloat[HeteroMixing module.]{
    \begin{minipage}[t]{0.31\linewidth}
    \centering
    \includegraphics[width=0.92\textwidth]{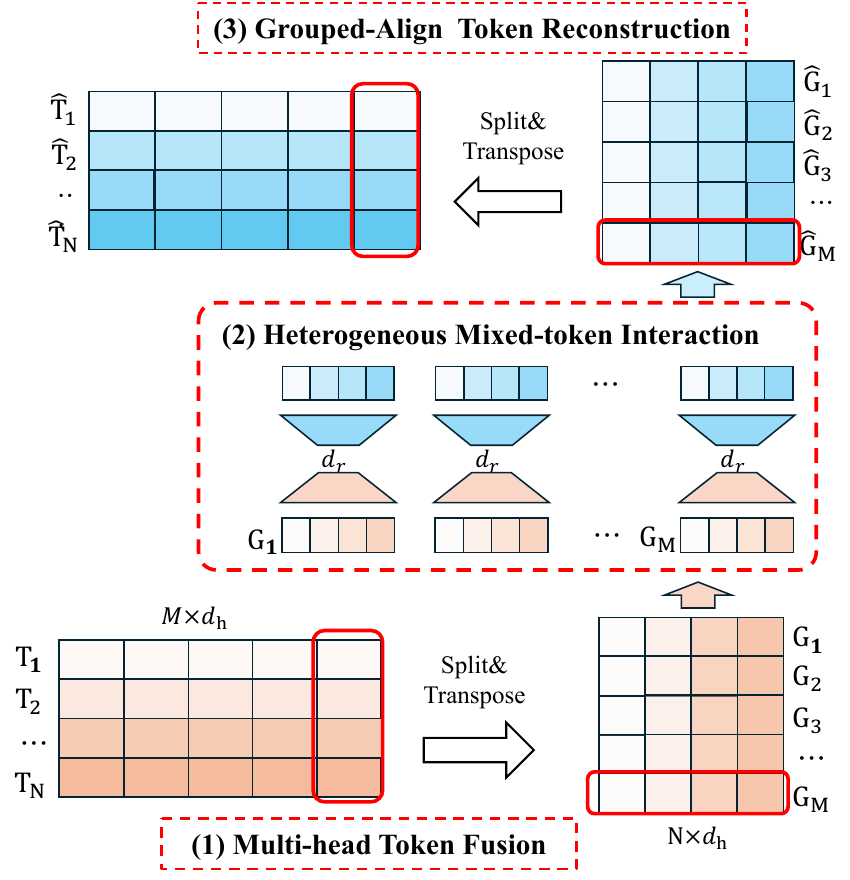}
    \end{minipage}
}
\centering
\caption{The overall architecture of HeMix with the feature embedding\& tokenization layer and the HeteroMixing module. }
\label{fig:hemix}
\end{figure*}

HeMix consists of three components: an embedding and tokenization layer, a feature interaction layer, and a prediction layer. \textbf{Embedding and tokenization layer.} Heterogeneous sparse features are mapped into dense embeddings via lookup tables and unified into $N$ dimension-aligned tokens. For non-sequential features, an \textit{interact-then-split} pipeline (Concat--MLP--Split) produces the NS tokens. For sequential features (the global historical and real-time behavior sequences), a \textit{Mixed Hetero Attention} mechanism extracts both context-aware and context-invariant user-interest tokens from each sequence. \textbf{Feature interaction layer.} It stacks $L$ HeteroMixer blocks to model high-order interactions among the $N$ semantic tokens. Each block replaces the computationally expensive Multi-Head Self-Attention~\cite{vaswani2017attention} with an efficient \textit{HeteroMixing} layer for heterogeneous multi-head token mixing, and the standard FFN with a \textit{Heterogeneous Feed-Forward Network} (HeteroFFN), followed by a residual connection and Layer Normalization. \textbf{Prediction layer.} It produces the final prediction from the interaction representations, tailored to a specific objective, e.g., CTR or CVR prediction.

\subsection{Input Layer and Feature Embedding} 
\label{sec:token}

\textbf{Input Features.} In a cascaded industrial recommender system, the recall stage retrieves a set of candidate items for a user $u$, and the ranking model estimates a score for each candidate item $i$. The ranking model consumes rich features that span categorical, continuous, and sequential behavior signals. To preserve feature heterogeneity and distinct structural semantics of each source, we organize the raw input of a user--item pair into five disjoint groups:
\begin{align}
\mathcal{S}_{u,i} = [
\underbrace{\mathbf{U}}_{\text{User Features}},
\underbrace{\mathbf{I}}_{\text{Item Features}},
\underbrace{\mathbf{C}}_{\text{Cross Features}},
\underbrace{\mathbf{RT\_S}}_{\text{RT Seq.}},
\underbrace{\mathbf{G\_S}}_{\text{Global Seq.}}
].
\end{align}
Here $\mathbf{U}$, $\mathbf{I}$, and $\mathbf{C}$ are \textit{non-sequential} (NS) features: $\mathbf{U}$ describes user profiles and contextual signals such as timestamps and geographic locations that reflect the user's situational state at request time, $\mathbf{I}$ describes attributes of the candidate item, and $\mathbf{C}$ carries user--item cross statistics. The remaining two groups are \textit{sequential} features, which record the user's historically interacted items together with their associated attributes. The \textit{global sequence} ($\mathbf{G\_S}$) covers the user's long-term interaction history, whereas the \textit{real-time sequence} ($\mathbf{RT\_S}$) contains only the interactions of the current day and is disjoint from the global sequence. This separation is deliberate: the two sequences differ substantially in both length and timeliness, and therefore call for distinct modeling strategies rather than a single unified treatment. Each group comprises multiple concrete fields, as illustrated in Figure~\ref{fig:hemix}(b).

\textbf{Feature Embedding.} For each input sample $\mathcal{S}_{u,i} \in \mathbb{S}$, sparse high-dimensional features are mapped into dense low-dimensional embeddings through group-specific lookup tables, where each feature group is assigned its own embedding dimension. The non-sequential features in $\mathbf{U}$, $\mathbf{I}$, and $\mathbf{C}$ are embedded and concatenated into an aggregated representation $\mathbf{E}_{U,I,C} \in \mathbb{R}^{d_{NS}}$, where $d_{NS}$ denotes their total dimensionality. For sequential features, each interacted item is represented by concatenating all of its side-information embeddings into a single dense vector. We denote the global sequence as $\mathbf{G} = [I_1, I_2, \ldots, I_{L_G}] \in \mathbb{R}^{L_G \times d_{I}}$ and the real-time sequence as $\mathbf{R} = [I_1, I_2, \ldots, I_{L_R}] \in \mathbb{R}^{L_R \times d_{I}}$, both arranged in reverse chronological order so that the most recent behavior comes first. Typically, $L_G$ is on the order of hundreds while $L_R$ is on the order of tens, which makes it prohibitively expensive to feed the raw sequences directly into the interaction layer thereby motivates the tokenization design in Section~\ref{sec:token2}.

\subsection{Feature Tokenization} 
\label{sec:token2}

To enable efficient parallel computation in later stages, embeddings of varying dimensions must be transformed into dimension-aligned vectors, called feature-tokens~\cite{zhu2025rankmixer}. Notably, we first generate the non-sequential tokens (referred to as NS tokens), and then compress thousands of sequence features to tens of representative sequence tokens based on the NS tokens with queries-mixed hetero attention.

\subsubsection{Non-sequential Features Tokenization.} 
In the previous stage, all non-sequential features have been concatenated into one embedding representation $\mathbf{E}_{U,I,C}$. A common approach is to apply AutoSplit to partition this concatenated vector directly. However, this strategy suffers from two limitations. \textit{First}, in large-scale recommender systems the concatenated vector spans hundreds of highly heterogeneous fields, so partitioning it at a fixed stride cuts across field boundaries and groups together whichever fields happen to be \textit{adjacent} in an engineering-defined layout. Such splitting disrupts the semantic boundaries among features, leading to a mixed feature space that impedes effective structured interaction in subsequent layers. \textit{Second}, a parameter-free partition enforces $N_{NS}\cdot d_T = d_{NS}$, which ties the token layout to the embedding-table configuration and prevents $N_{NS}$ and $d_T$ from being chosen independently.

To address these issues, we propose an ``interact-then-split'' paradigm. Specifically, we first model high-order bit-wise feature interactions using an implicit interaction function $\mathcal{F}_{NS}(\cdot):\mathbb{R}^{d_{NS}}\!\to\!\mathbb{R}^{N_{NS}\cdot d_T}$, which can be a multi-layer perceptron (MLP)~\cite{lian2018xdeepfm} with nonlinear activations (e.g., ReLU). By setting the output dimension of the final MLP layer to $N_{NS}\cdot d_T$, we ensure its output can be evenly partitioned into sub-vectors that serve as tokens for NS features:
\begin{align}
    \mathbb{T}_{NS} = Split(\mathcal{F}_{NS}(\mathbf{E}_{U,I,C}), split\_size=d_T) \in \mathbb{R}^{N_{NS} \times d_T},
\end{align}
where $Split(\cdot, split\_size)$ splits the input tensor into tensors of equal dimension $d_T$, and $N_{NS}$ and $d_T$ are hyper-parameters. Two remarks are in order. First, since the output width of $\mathcal{F}_{NS}$ is chosen by construction, $N_{NS}$ and $d_T$ are decoupled from $d_{NS}$, which is what makes the orthogonal scaling along $N$ and $d_T$ in Section~\ref{sec:scaling} available at all. Second, writing the partition as $\mathbb{T}_{NS}=\mathbf{P}\,\mathbf{E}_{U,I,C}$ with a fixed binary selection matrix $\mathbf{P}$, AutoSplit is recovered as the degenerate case in which the first layer of $\mathcal{F}_{NS}$ is frozen to $\mathbf{P}$; the learned tokenizer is therefore at least as expressive.

\subsubsection{Query-mixed Interest Extraction for Sequence Tokenization.}
As introduced earlier, we utilize two types of sequential features to jointly model users' global- and real-time interest preferences: the global sequence, consisting of the user's historical interactions, encodes long-term interests; the real-time sequence, comprising interactions from the current day, captures the user's current spatio-temporal context and transient intent. Guided by the NS tokens $\mathbb{T}_{NS}$, we employ a heterogeneous attention mechanism to dynamically extract context-aware user interest representations from these sequences. In this process, we partition the NS tokens into two groups in a $4\!:\!1$ ratio, i.e., $\mathbb{T}_{\text{NS}} = [\mathbb{T}_{\text{NS}}^{G}, \mathbb{T}_{\text{NS}}^{R}]$, with the larger portion $\mathbb{T}_{\text{NS}}^{G}$ used for attending to the global sequence and the smaller part $\mathbb{T}_{\text{NS}}^{R}$ for the real-time sequence.

Furthermore, inspired by the Q-Former in BLIP-2~\cite{li2023blip}, we additionally initialize two groups of learnable query embeddings $\mathbf{Q}=[\mathbf{Q}^G,\mathbf{Q}^R] \in \mathbb{R}^{N_{NS} \times d_T}$, shaped identically to $\mathbb{T}_{\text{NS}}$, for learning \textit{context-invariant} sequence features that represent inherent behavioral patterns shared across a large population of users. The two groups interact separately with the global and real-time sequences. Crucially, the two query types are complementary rather than redundant: a dynamic query is a function of the current request, so it can only surface behaviors that correlate with the present context, which systematically down-weights patterns that are predictive yet uncorrelated with the candidate; a fixed query is request-independent and therefore recovers exactly such patterns. Because what a fixed query extracts depends on the behavior sequence alone, it also provides a lower-variance signal for users whose sequences are short or noisy, where candidate-conditioned attention is easily dominated by spurious matches. This improves the model's understanding of sequential patterns and its overall stability.

\begin{figure}[t]
    \setlength{\abovecaptionskip}{0.2cm}
    \setlength{\belowcaptionskip}{-0.4cm}
    \centering
    \includegraphics[width=0.70\linewidth]{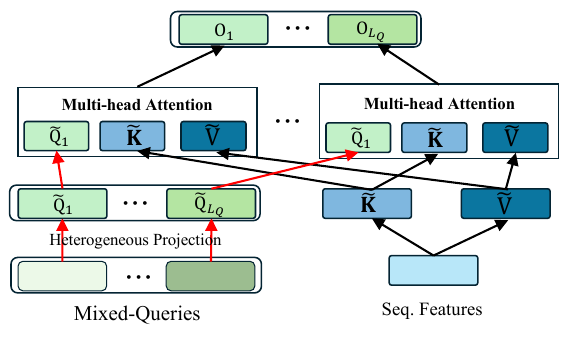}
    \caption{The structure of Mixed Hetero Attention.}
    \label{fig:mha}
\end{figure}

Take the global sequence as an example (Figure~\ref{fig:mha}). We form the mixed global queries $\mathbb{Q} = [\mathbb{T}_{\text{NS}}^{G}, \mathbf{Q}^G] \in \mathbb{R}^{L_{Q} \times d_T}$, where $L_Q$ is twice the size of $\mathbb{T}_{\text{NS}}^{G}$, and take the global sequence matrix $\mathbf{G} \in \mathbb{R}^{L_G \times d_{I}}$ as both Key and Value, i.e., $\mathbf{K}=\mathbf{V}=\mathbf{G}$. In standard Transformers, the attention score is computed as
\begin{align}
\mathbf{Att}(\mathbb{Q}, \mathbf{K}) = \operatorname{softmax}\big((\mathbb{Q} \mathbf{W}_Q) (\mathbf{K} \mathbf{W}_K)^{\!\top}/{\sqrt{d}}\big),
\end{align}
which applies a shared $\mathbf{W}_Q$ and $\mathbf{W}_K$ to all queries. Expanding the logits as $\mathbb{Q}^{(l)}\,(\mathbf{W}_Q \mathbf{W}_K^{\!\top})\,\mathbf{K}^{\!\top}$ reveals the limitation: every query scores behaviors through one and the same bilinear form $\mathbf{W}_Q \mathbf{W}_K^{\!\top}$, so different queries differ only in \textit{where} they sit within a single similarity metric, not in \textit{what} similarity they measure. This is ill-suited to our setting, where the queries are deliberately heterogeneous---some are candidate-conditioned, others are population-level priors. Hence, we assign a dedicated projection matrix to each query vector,
\begin{align}
    [\tilde{\mathbb{Q}}^{(1)},...,\tilde{\mathbb{Q}}^{(L_{Q})}] = [\mathbb{Q}^{(1)} \mathbf{W}_Q^{(1)},...,\mathbb{Q}^{(L_{Q})} \mathbf{W}_Q^{(L_{Q})}],
\end{align}
where $\mathbb{Q}^{(l)}$ denotes the $l$-th query and $[\mathbf{W}_Q^{(1)};...;\mathbf{W}_Q^{(L_{Q})}] \in \mathbb{R}^{L_{Q} \times d_T \times d}$ are heterogeneous query projections. Each query thereby induces its own bilinear form $\mathbf{W}_Q^{(l)}\mathbf{W}_K^{\!\top}$, i.e., its own metric over behaviors. Meanwhile, to reduce parameters, we retain shared projections $\mathbf{W}_K, \mathbf{W}_V \in \mathbb{R}^{d_I \times d}$ for key and value:
\begin{align}
    \tilde{\mathbf{K}} = \mathbf{K} \mathbf{W}_K, \quad
    \tilde{\mathbf{V}} = \mathbf{V} \mathbf{W}_V.
\end{align}
Sharing $\mathbf{W}_K$ is not merely an efficiency measure: it constrains all $L_Q$ bilinear forms to factor through a common key subspace, which regularizes the family of learned metrics while still allowing each query to specialize.

Then, we compute the heterogeneous attention scores and the corresponding output for each query:
\begin{align}
\mathbf{Att}^{(l)} &= \operatorname{softmax}\big((\tilde{\mathbb{Q}}^{(l)} \tilde{\mathbf{K}}^{\!\top})/{\sqrt{d}}\big), \\
\mathbf{O}^{(l)} &= \mathbf{Att}^{(l)} \tilde{\mathbf{V}} \in \mathbb{R}^{1 \times d}.
\end{align}
Stacking the $L_Q$ per-query outputs and applying an output projection $\mathbf{W}_O \in \mathbb{R}^{d \times d_T}$ yields
\begin{equation}
\mathbf{O}^{G} = \big[ \mathbf{O}^{(1)}; \dots; \mathbf{O}^{(L_Q)} \big] \mathbf{W}_O \in \mathbb{R}^{L_Q \times d_T}.
\end{equation}
In practice, each heterogeneous projection can be further split into multiple heads to obtain better performance. Note that this cross-attention formulation costs $\mathcal{O}(L_Q L_G d)$, i.e., it is \textit{linear} in the sequence length, in contrast to the $\mathcal{O}(L_G^2 d)$ cost incurred by flattening-based tokenizers that run self-attention over the raw behavior sequence; since $L_Q \ll L_G$, the sequence is compressed at a small fraction of that cost. The resulting $\mathbf{O}^G$ serves as the global sequence tokens, encoding both context-aware and context-invariant user interests through heterogeneous query-specific projections.

Similarly, for the real-time sequence $\mathbf{R}$, we apply the same pipeline with mixed queries $\mathbb{Q}_{RT} = [\mathbb{T}_{\text{NS}}^{R}, \mathbf{Q}^R]$:
\begin{align}
    \mathbf{O}^{R} = \textit{Mixed Hetero Attention}(\mathbb{Q}_{RT}, \mathbf{R}, \mathbf{R}).
\end{align}
The output $\mathbf{O}^{R}$ is taken as the real-time sequence tokens.

\textbf{Token Assembly.} The complete sequential tokens are obtained by combining $\mathbf{O}^{G}$ and $\mathbf{O}^{R}$, denoted as $\mathbb{T}_{S} = [\mathbf{O}^{G},\mathbf{O}^{R}] \in \mathbb{R}^{N_S \times d_T}$. Since each sequence is attended by both a dynamic and a fixed query group of equal size, the number of sequence tokens is exactly twice that of the NS tokens, i.e., $N_S = 2 N_{NS}$. Combining with the NS tokens yields the full token set for downstream interaction:
\begin{align}
    \mathbb{T} = [\mathbb{T}_{S},\mathbb{T}_{NS}] \in \mathbb{R}^{N \times d_T},
\end{align}
where $N = N_S + N_{NS}$ is the total number of tokens. $\mathbb{T}$ is then fed into the interaction layer with $L$ successive HeteroMixer Blocks.

\subsection{Interaction Layer with HeteroMixer Block}
\label{sec:hetero}

While the tokens from the tokenization layer encode preliminary feature signals and user interests, their representational capacity remains insufficient for industrial-scale ranking with highly dense features. To capture high-order dependencies among heterogeneous tokens, we stack $L$ HeteroMixer blocks in a Transformer-style architecture. Within each block, we replace the standard Multi-Head Self-Attention with our proposed \textit{HeteroMixing} layer and the standard FFN with a \textit{Heterogeneous FFN} (HeteroFFN), while retaining the residual and Layer Normalization (LN) structures. Formally, the $l$-th block updates the token representations as
\begin{align}
     \mathbf{S}_{l} &= LN\big(HeteroMixing(\mathbb{T}_{l-1}) + \mathbb{T}_{l-1}\big), \label{eq:blk1}\\
     \mathbb{T}_{l} &= LN\big(HeteroFFN(\mathbf{S}_{l}) + \mathbf{S}_{l}\big), \label{eq:blk2}
\end{align}
where $\mathbb{T}_{0} = \mathbb{T} = [\mathbf{T}_1, \mathbf{T}_2, \dots, \mathbf{T}_{N}] \in \mathbb{R}^{N \times d_T}$ is the input token set produced by the tokenization layer, $\mathbf{S}_l$ is the intermediate state, and $\mathbb{T}_l$ denotes the output of the $l$-th block. $LN(\cdot)$ is the Layer Normalization function. The two modules are not interchangeable but \textit{dual} to each other: HeteroMixing mixes information across tokens, whereas HeteroFFN mixes information across channels within a token. We detail them below and then analyze their interplay in Section~\ref{sec:hetero_disc}, omitting the block index $l$ for clarity.

\subsubsection{HeteroMixing Layer}
To enable efficient interaction over heterogeneous tokens, we propose the \textit{HeteroMixing} layer (Figure~\ref{fig:hemix}(c)), which replaces the $\mathcal{O}(N^2 d_T)$ pairwise self-attention operation with a \textit{fuse--interact--reconstruct} pipeline built on structured tensor reshaping. It comprises three stages: (i) \textit{Multi-Head Token Fusion} regroups the tokens into $M$ mixed tokens; (ii) \textit{Heterogeneous Mixed-Token Interaction} performs lightweight non-linear mixing across all tokens within each head; and (iii) \textit{Group-Aligned Reconstruction} restores the original token layout.

\textbf{Multi-Head Token Fusion.}
We split each token $\mathbf{T}_n \in \mathbb{R}^{d_T}$ into $M$ sub-tokens of dimension $d_h = d_T / M$, and concatenate sub-tokens of the same head across all $N$ tokens to form a \textit{mixed token}:
\begin{align}
    [\mathbf{T}^{(1)}_n,...,\mathbf{T}^{(M)}_n ] &= Split(\mathbf{T}_n, split\_size=d_h), \\
    \mathbf{G}_m &= concat([\mathbf{T}^{(m)}_1,\mathbf{T}^{(m)}_2,...,\mathbf{T}^{(m)}_N ]) \in \mathbb{R}^{N\cdot d_h}, \\
    \mathbb{G} &= [\mathbf{G}_{1}, \mathbf{G}_{2}, ..., \mathbf{G}_{M}] \in \mathbb{R}^{M \times N\cdot d_h},
\end{align}
where $\mathbf{T}^{(m)}_n$ denotes the $m$-th sub-token of the $n$-th token. Each mixed token $\mathbf{G}_m$ thus aggregates a distinct semantic slice across all $N$ heterogeneous tokens. Note that this stage is a bijective re-indexing of the $N d_T$ coordinates rather than a compression, so it is information-preserving and parameter-free.

\textbf{Heterogeneous Mixed-Token Interaction.}
After fusion, each mixed token $\mathbf{G}_m$ encodes a head-specific cross-token view, e.g., user, item, and sequence features collectively projected onto the $m$-th sub-space. To capture non-linear high-order dependencies within each view, we apply an independent two-layer MLP with low-rank decomposition to every mixed token in parallel:
\begin{align}
    \hat{\mathbf{G}}_{m} &= \mathbf{W}^{R}_m \, \mathrm{ReLU}(\mathbf{W}^{L}_m \mathbf{G}_{m}), \\
    \hat{\mathbb{G}} &= [\hat{\mathbf{G}}_{1}, \hat{\mathbf{G}}_{2}, ..., \hat{\mathbf{G}}_{M}] \in \mathbb{R}^{M \times N\cdot d_h},
\end{align}
where $\mathbf{W}^{L}_m \in \mathbb{R}^{d_r \times N \cdot d_h}$ and $\mathbf{W}^{R}_m \in \mathbb{R}^{N \cdot d_h \times d_r}$, with $d_r \ll N \cdot d_h$. Crucially, the $M$ mixed tokens are processed by \textit{head-specific} weights, allowing each semantic group to learn its own interaction pattern, which is the source of \textit{heterogeneity} in our design.

\textbf{Group-Aligned Reconstruction.}
Finally, the refined mixed-token matrix $\hat{\mathbb{G}} \in \mathbb{R}^{M \times N\cdot d_h}$ is inversely reshaped back to the original layout according to the initial group assignments, yielding $\hat{\mathbb{T}} = HeteroMixing(\mathbb{T}) \in \mathbb{R}^{N \times d_T}$. This step is exactly the structural inverse of the fusion stage and introduces no parameters. Each reconstructed token thereby carries both the local semantics of its own sub-spaces and the global context aggregated from all other tokens, and is subsequently fused with the block input through the residual connection and Layer Normalization of Eq.~(\ref{eq:blk1}).

\textbf{Complexity Analysis.}
The two design choices, i.e., low-rank factorization and multi-head fusion, in the interaction stage admit a precise accounting.
\textit{First, low-rank factorization is necessary rather than merely convenient.} A dense per-head mixing matrix $\mathbf{W}_m \in \mathbb{R}^{N d_h \times N d_h}$ would require $\sum_{m=1}^{M}(N d_h)^2 = N^2 d_T d_h$ operations, which is \textit{worse} than the $\mathcal{O}(N^2 d_T)$ cost of pairwise self-attention by a factor of $d_h$. The rank-$d_r$ factorization reduces this to $\sum_{m=1}^{M} 2 N d_h d_r = 2 N d_T d_r = \mathcal{O}(N d_T d_r)$, i.e., \textit{linear} in the number of tokens. Low-rank is thus precisely what turns the fuse--interact--reconstruct pipeline into a viable substitute for attention rather than a more expensive one. It additionally imposes a useful inductive bias: each head must summarize its $N$-token slice into $d_r$ shared factors before redistributing them, which favors a small number of global interaction patterns over $N^2$ independently parameterized pairwise couplings.
\textit{Second, multi-head fusion multiplies the mixing rank at zero cost.} Since $\sum_{m} 2 N d_h d_r = 2 N d_T d_r$ holds for any $M$, both the parameter count and the FLOP count of this stage are \textit{independent} of $M$. Yet the induced operators differ substantially: with $M\!=\!1$ the entire $N d_T$-dimensional token matrix is routed through a single rank-$d_r$ bottleneck, whereas $M$ heads provide $M$ parallel bottlenecks whose composite map attains rank up to $M d_r$. Multi-head fusion therefore increases the expressive rank of the mixing operator by a factor of $M$ at identical cost. The price is that the $M$ sub-spaces are never mixed with one another, which is exactly the role delegated to HeteroFFN.

\subsubsection{Heterogeneous Feed-Forward Network}

We further replace the shared FFN in vanilla Transformers~\cite{vaswani2017attention} with a \textit{Heterogeneous FFN} (HeteroFFN), in which each token is processed by its own dedicated two-layer MLP. This design is motivated by the heterogeneity of our token set: different tokens originate from semantically distinct sources, namely the non-sequential feature groups ($\mathbf{U}$, $\mathbf{I}$, $\mathbf{C}$) and the interests extracted from the behavior sequences ($\mathbf{RT\_S}$, $\mathbf{G\_S}$) by the dynamic and fixed queries. They therefore benefit from token-specific non-linear transformations rather than from a single shared projection. Formally, given the $i$-th token $\mathbf{S}^{(i)}$ of the HeteroMixing output $\mathbf{S}$, we compute
\begin{align}
    \hat{\mathbf{S}}^{(i)} = \mathrm{ReLU}\!\big(\mathbf{S}^{(i)} \mathbf{W}^{(i)}_1 + \mathbf{b}^{(i)}_1\big)\,\mathbf{W}^{(i)}_2 + \mathbf{b}^{(i)}_2,
\end{align}
where $\mathbf{W}^{(i)}_1 \in \mathbb{R}^{d_T \times k d_T}$, $\mathbf{W}^{(i)}_2 \in \mathbb{R}^{k d_T \times d_T}$, $\mathbf{b}^{(i)}_1 \in \mathbb{R}^{k d_T}$ and $\mathbf{b}^{(i)}_2 \in \mathbb{R}^{d_T}$ are token-specific parameters, and $k$ is the hidden expansion ratio (set to $4$ by default). Stacking the outputs of all $N$ tokens gives
\begin{align}
    \hat{\mathbf{S}} = \big[\hat{\mathbf{S}}^{(1)}, \dots, \hat{\mathbf{S}}^{(N)}\big] = HeteroFFN\big(\mathbf{S}^{(1)},...,\mathbf{S}^{(N)}\big).
\end{align}
Although HeteroFFN holds $N\times$ more parameters than a shared FFN, all $N$ projections are executed in parallel via batched matrix multiplication and incur negligible latency overhead in practice, for the reason quantified below. The output $\hat{\mathbf{S}}$ is then fused with $\mathbf{S}$ via a residual connection and Layer Normalization following Eq.~(\ref{eq:blk2}), producing the block output $\mathbb{T}_{l}$. After stacking $L$ HeteroMixer blocks, the final representations $\mathbb{T}_L$ encode high-order, heterogeneity-aware token interactions and are passed to the prediction layer.

\subsubsection{Discussion}
\label{sec:hetero_disc}
Two block-level properties explain why the two modules are jointly necessary rather than interchangeable.

\textit{(i) They mix along complementary axes.} By the construction of $\mathbf{G}_m$, the output coordinate (token $n$, sub-space $m$) of HeteroMixing depends on \textit{all} $N$ tokens but only within sub-space $m$; HeteroFFN is the exact dual, mixing all $d_T$ channels of a token but never crossing tokens. Their composition within one block therefore attains a \textit{full} receptive field over all $N d_T$ coordinates, while each factor alone costs $\mathcal{O}(N)$. This factorization into a token axis and a channel axis reconciles global reach with linear cost; removing either module leaves one axis unmixed.

\textit{(ii) Heterogeneity is free in FLOPs.} Sharing $\{\mathbf{W}^{L}_m, \mathbf{W}^{R}_m\}$ across heads, or sharing a single FFN across tokens, would reduce the parameter count by a factor of $M$ and $N$ respectively, yet leave the arithmetic cost unchanged at $2 N d_T d_r$ and $2 N k d_T^2$ multiply--accumulate operations: the same matrix products must be evaluated once per head and once per token either way. Since a shared parameterization is exactly the special case in which all heads (or all tokens) are tied, the heterogeneous variant spans a strictly larger function class at identical arithmetic cost, and Table~\ref{exp:abl} confirms that this additional capacity is indeed realized as accuracy. The extra weights do increase memory traffic, but a single ranking request scores hundreds of candidates in one batch, so each weight is loaded once and reused across the whole batch, which keeps both modules compute-bound in practice. Finally, the residual paths guarantee a safe fallback: setting $\mathbf{W}^{R}_m\!\to\!\mathbf{0}$ deactivates the mixing branch and recovers the normalized identity $LN(\mathbb{T})$, so the block is never forced to apply an unhelpful transform and is at least as expressive as a parameter-free reshape.

Overall, HeteroMixer occupies the middle ground between pairwise self-attention~\cite{vaswani2017attention}, which models all token pairs at a prohibitive $\mathcal{O}(N^2 d_T)$ cost, and parameter-free token reshaping~\cite{zhu2025rankmixer}, which is a linear re-indexing that adds no capacity of its own: it retains the efficiency of structural reshaping while reinjecting non-linear, heterogeneity-aware interaction at linear cost.

\subsection{Prediction Layer}
\label{sec:prediction}
The final token representations $\mathbb{T}_L$ are aggregated by mean pooling into a unified vector $\mathbf{o}_{\text{out}} \in \mathbb{R}^{d_T}$, from which the predicted click probability of the $i$-th sample is obtained as
\begin{align}
    \hat{y}_i = \sigma\!\big(f(\mathbf{o}_{\text{out}})\big),
\end{align}
where $f(\cdot)$ denotes the prediction head, instantiated as an MLP in our implementation, and $\sigma(\cdot)$ is the sigmoid function. The model is optimized end-to-end by minimizing the binary cross-entropy loss over the training set $\mathcal{D}$:
\begin{align}
    \mathcal{L}_{\text{CTR}} = -\frac{1}{|\mathcal{D}|}\sum_{i\in\mathcal{D}} \big[\, y_i \log \hat{y}_i + (1-y_i)\log(1-\hat{y}_i)\,\big],
\end{align}
where $y_i \in \{0,1\}$ is the true click label. The CVR objective $\mathcal{L}_{\text{CVR}}$ is defined analogously, with $y_i$ replaced by the conversion label and a separate prediction head applied to the same shared representation.

\subsection{Scaling Up HeMix}
\label{sec:scaling}
A key advantage of HeMix is that its capacity can be scaled along four independent hyperparameters: the number of tokens $N$, the token dimension $d_T$, the HeteroFFN expansion ratio $k$, and the block depth $L$. The parameter cost is dominated by the token-specific projections in HeteroFFN, so the total parameter count is well approximated by
\begin{align}
    \#\text{Params} \;\approx\; 2\,k\,L\,N\,d_T^2,
\end{align}
which grows linearly in $k$, $L$ and $N$, and quadratically in $d_T$. HeteroMixing, by contrast, contributes only $2\,L\,N\,d_T\,d_r$ parameters under the low-rank decomposition; since $d_r \ll k\,d_T$ in practice, this term is negligible by comparison. Parameter scaling (driven by HeteroFFN) is therefore decoupled from interaction modeling (handled by the lightweight HeteroMixing), which allows HeMix to grow in capacity without incurring the quadratic cost in $N$ that pairwise attention would impose. Section~\ref{sec:scaling_exp} verifies empirically that this design yields a log-linear scaling law over more than one order of magnitude in model size.

\section{Experiments}

\subsection{Experiment Settings}
\subsubsection{Dataset and Environment.} The offline experiments are conducted on the AMAP platform\footnote{\url{https://amap.com}}, a large-scale navigation-oriented LBS recommender system. The dataset is constructed from three consecutive months of real-world online logs and user behavior records. To ensure privacy compliance, all user-related information is anonymized via cryptographic hashing before being used for model training and evaluation. The overall dataset contains over 4.4 billion samples, including more than 350 million clicked instances. Detailed statistics are summarized in Table \ref{tab:dataset}.  This dataset captures the core challenges of industrial recommendation: extreme scale, high sparsity, and complex cross-field interactions.

\subsubsection{Evaluation Metrics}
We adopt AUC and Logloss (LL) to evaluate model performance. A slightly higher AUC or a lower Logloss at a 0.001-level is considered statistically significant ~\cite{wang2020dcn,liu2019feature}, as even small CTR gains can translate to huge revenue increases in large-scale applications~\cite{fat2025, zhang2024wukong}. The \textit{Rela.Imp} (Relative Improvement) quantifies performance gains relative to a base model:
\begin{align}
\textstyle
\textit{Rela.Imp}=\left(\frac{\text{AUC}(\text {measure model})-0.5}{\text{AUC}(\text{base model})-0.5}-1\right) \times 100 \% .
\end{align} 

For robustness, each experiment is repeated four times, with averaged results reported.  The significance is verified via the two-tailed unpaired t-test~\cite{liu2019feature, li2023ctrl}, and all results satisfy the significance test: $ p< 0.05$ compared to the best baseline.

\begin{table}[t]
    \centering
    \caption{Dataset Statistics. A.L(G) and A.L(RT) are the average lengths for the global and real-time sequence respectively.}
    \scalebox{0.9}{
    \begin{tabular}{ccccccc}
        \hline
    \hline
    Dataset & \#Users & \#Items  & A.L(G) & A.L(RT) & \#Sample & \#Clicked   \\
        \hline
        AMAP&2M+&2.5M+ &332 & 10 &4.4B &350M \\ 
    \hline
    \hline
    \end{tabular}
    }
    \label{tab:dataset}
\end{table}

\begin{table*}
    \centering
    \caption{Offline CTR/CVR performance (AUC, LogLoss) and efficiency (\#Params, GFLOPs) at two model scales (${\sim}100$M and ${\sim}1500$M). \textbf{Bold}: best per scale; \underline{underline}: second-best. $\star$: statistically significant over the second-best ($p<0.05$, two-sided $t$-test). GFLOPs measured via TensorFlow profiling. Rela.Imp denotes relative improvement over DLRM.}
    \setlength{\tabcolsep}{2.0mm}
    \scalebox{0.99}{
    \begin{tabular}{c|cccc|cccc|cc} 
    \hline
    \hline
    \multirow{2}{*}{Methods} & \multicolumn{4}{c|}{CTR}   & \multicolumn{4}{c|}{CVR}  & \multicolumn{2}{c}{Efficiency}  \\ 
    \cline{2-11}
  & AUC$\uparrow$  & Rela.Imp   & Logloss$\downarrow$ &Rela.Imp& AUC$\uparrow$    & Rela.Imp   & Logloss$\downarrow$ &Rela.Imp & \#Param (M) & GFLOPs \\ 
    \hline
    DLRM    & 0.7312 & -      & 0.2563  & -       & 0.7855 & -      & 0.1760  & -       & 13 & 0.35   \\
    DLRM-L       & 0.7315 & 0.13\% & 0.2558  & -0.20\% & 0.7859 & 0.14\% & 0.1751  & -0.51\% & 91 & 2.60    \\
    DCNv2  & 0.7320 & 0.35\% & 0.2547  & -0.62\% & 0.7868 & 0.46\% & 0.1743  & -0.97\% & 47 & 2.41   \\
    AutoInt+& 0.7322 & 0.43\% & 0.2545  & -0.70\% & 0.7871 & 0.56\% & 0.1741  & -1.08\% & 111& 3.23   \\ 
    \hline
    HiFormer        & 0.7326 & 0.61\% & 0.2541  & -0.86\% & 0.7877 & 0.77\% & 0.1730  & -1.70\% & 103& 1.93   \\
    MTGR   & 0.7339 & 1.17\% & 0.2520  & -1.68\% & 0.7893 & 1.33\% & 0.1713  & -2.67\% & 79 & 1.49   \\
    RankMixer       & 0.7341 & 1.25\% & 0.2516  & -1.83\% & 0.7894 & 1.37\% & 0.1711  & -2.78\% & 110& 2.07   \\ 
    HyFormer	&0.7343	&1.34\%	&\underline{0.2511}	&-2.03\%&	\underline{0.7898}&	1.51\%	&\underline{0.1707}	&-3.01\%&	105	&2.51 \\
    OneTrans	&\underline{0.7344}	&1.38\%	&0.2513	&-1.95\%	&0.7896	&1.44\%	&0.1710	&-2.84\%	&97	&2.32 \\
    \textbf{HeMix(Ours)}        & $\textbf{0.7350}^{\star}$ & \textbf{1.64\%} & $\textbf{0.2495}^{\star}$ & \textbf{-2.65\%} & $\textbf{0.7902}^{\star}$ & \textbf{1.63\%} & $\textbf{0.1696}^{\star}$  & \textbf{-3.64\%} & 101& 1.89   \\
    \hline
    RankMixer-L       & 0.7357  & 1.95\%  & 0.2492   & -2.77\%  & 0.7912  & 2.14\%  & 0.1680  & -4.55\%  & 1532& 30.18   \\ 
    HyFormer-L	&0.7356	&1.90\%	&0.2496	&-2.61\%	&0.7910&	1.93\%	&0.1683	&-4.37\%	&1382	&30.63 \\
    OneTrans-L	&\underline{0.7359}	&2.03\%	&\underline{0.2490}	&-2.85\%	&\underline{0.7914}	&2.07\%	&\underline{0.1678}	&-4.66\%	&1354	&34.49 \\
    \textbf{HeMix-L(Ours)}        & $\textbf{0.7366}^{\star}$ &\textbf{2.34\%}& $\textbf{0.2476}^{\star}$  & \textbf{-3.39\%} &$\textbf{0.7923}^{\star}$ & \textbf{2.38\% }& $\textbf{0.1665}^{\star}$  & \textbf{-5.40\%} & 1484        & 27.58  \\
    \hline
    \hline
    \end{tabular}
}
\label{exp:overall}
\end{table*}

\subsubsection{Baselines.}
We compare HeMix against representative state-of-the-art methods, categorized by modeling paradigm:
\begin{itemize}[leftmargin=*, topsep=2pt, itemsep=1pt]
\item \textbf{Traditional Interaction Models}: DLRM~\cite{naumov2019deep}, DCNv2~\cite{wang2020dcn} and AutoInt+~\cite{song2019autoint}, which emphasize structured or learned pairwise feature crosses over a flat embedding table. 
\item \textbf{Scaling-Oriented Architectures}: HiFormer~\cite{gui2023hiformer}, RankMixer~\cite{zhu2025rankmixer}, MTGR~\cite{han2025mtgr},  HyFormer~\cite{huang2026hyformer} and OneTrans~\cite{zhang2025onetrans}, which restructure the ranking backbone into stacked interaction blocks over tokenized features, representing recent advances in scalable model design.
\end{itemize}
For a fair comparison, all methods are aligned to two matched capacity regimes, ${\sim}100$M and ${\sim}1500$M parameters, by adjusting width or depth while preserving their core architectural constraints. Accordingly, we instantiate two variants of HeMix: HeMix (${\sim}100$M) and HeMix-L (${\sim}1500$M).

\subsubsection{Implementation Details.}
All experiments are conducted on a distributed cluster of 100 GPUs and implemented in TensorFlow. For fair comparison, all offline models are trained from scratch with identical hyperparameters (batch size $256$, learning rate $1\mathrm{e}{-4}$), and share the same optimizer and data preprocessing pipeline. By default, the token set consists of $N=60$ tokens of dimension $d_T=256$, comprising $N_{NS}=20$ NS tokens and $N_S=40$ sequence tokens. Following the $4\!:\!1$ split in Section~\ref{sec:token2}, the global and real-time sequences receive $16$ and $4$ dynamic queries plus an equal number of fixed queries, yielding $32$ and $8$ tokens respectively. The HeteroFFN expansion ratio is $k=4$, while the block depth $L$, the number of heads $M$, and the low-rank dimension $d_r$ are tuned per model scale. For online experiments, we adopt a warm start: sparse embedding tables are initialized from the production model, whereas the dense tokenization and interaction layers are randomly initialized, which stabilizes training while letting the new modules adapt to the serving distribution.

\subsection{Overall Comparison}

To demonstrate the effectiveness of HeMix, we conduct comprehensive offline experiments on the large-scale AMAP dataset over two industrial tasks: CTR and CVR prediction. We compare HeMix against state-of-the-art ranking models at two matched scales, i.e., ${\sim}100$M and ${\sim}1500$M parameters. Notably, following~\cite{zhu2025rankmixer}, all models within the same scale are aligned to comparable parameter counts and GFLOPs, so that the comparison reflects the intrinsic capability of each architecture rather than differences in computational budget. The prediction accuracy and efficiency results are jointly reported in Table~\ref{exp:overall}, from which we draw the following observations:

\textit{First, scaling-oriented architectures consistently outperform traditional feature-interaction models.} Even the weakest scaling-oriented method, HiFormer ($0.7326$ CTR-AUC), surpasses the traditional baseline, AutoInt+ ($0.7322$ CTR-AUC), at comparable parameter counts. HeMix further widens this advantage to $+1.21\%$ relative CTR-AUC over AutoInt+. The poor scaling of DLRM-L, which increases parameters by $7\times$ yet yields only $+0.0003$ CTR-AUC, confirms brute-force embedding expansion is insufficient, and effective performance gains require architectural support for token-level interaction.

\textit{Second, HeMix outperforms contemporary scaling-oriented baselines while being more efficient.} At ${\sim}100$M, HeMix leads all baselines in both CTR-AUC ($0.7350$) and CVR-AUC ($0.7902$), outperforming OneTrans ($0.7344$ / $0.7896$), RankMixer ($0.7341$ / $0.7894$), and HyFormer ($0.7343$ / $0.7898$) by statistically significant margins ($p<0.05$). This advantage is rooted in HeMix's two core designs. First, the Query-Mixed Interest Extraction module explicitly disentangles context-aware and context-invariant user interests from both global and real-time sequences, whereas OneTrans flattens heterogeneous behaviors into a single homogeneous token sequence and RankMixer compresses them into overly compact representations, losing fine-grained intent signals. Second, the HeteroMixer Block replaces the quadratic pairwise attention of OneTrans and HyFormer with $\mathcal{O}(N)$ heterogeneous token mixing, while injecting learnable non-linear interactions absent in RankMixer's parameter-free token reshuffling. As a result, HeMix achieves the best accuracy with the lowest computational cost ($1.89$ GFLOPs) among the top-performing ${\sim}100$M models. The same pattern holds at ${\sim}1500$M: HeMix-L ($0.7366$ / $0.7923$) outperforms OneTrans-L, RankMixer-L, and HyFormer-L while requiring $20\%$ fewer GFLOPs than OneTrans-L ($27.58$ vs. $34.49$).

\textit{Third, HeMix scales favorably with capacity.} From ${\sim}100$M to ${\sim}1500$M, HeMix gains $+0.0016$ CTR-AUC, matching RankMixer and exceeding OneTrans ($+0.0015$) and HyFormer ($+0.0013$). Because HeMix decouples parameter scaling (driven by HeteroFFN) from interaction modeling (handled by lightweight HeteroMixing), it avoids the $\mathcal{O}(N^2)$ attention bottleneck that constrains Transformer-based competitors such as OneTrans and HyFormer. Combined with its highest starting point, this orthogonal scaling property makes HeMix-L the strongest model at any compute budget shown.

\subsection{Ablation Study}

To verify the effectiveness of each design in HeMix, we conduct ablations on its two pillars: the \textit{tokenization layer} (C1) and the \textit{interaction layer} (C2). All variants are instantiated at the ${\sim}100$M scale with identical training settings:


\begin{table}
\centering
\caption{Ablation study of tokenization strategy.}
\label{exp:abl_token}
    \scalebox{0.95}{
    \begin{tabular}{c|cccc} 
    \hline
    \hline
    \multirow{2}{*}{Variants} & \multicolumn{2}{c}{CTR} & \multicolumn{2}{c}{CVR}  \\ 
    \cline{2-5}
    & AUC    & Logloss        & AUC    & Logloss\\ 
    \hline
    HeMix      & \textbf{0.7350} & \textbf{0.2495}& \textbf{0.7902}& \textbf{0.1696} \\
    w/o Fixed Query  & 0.7345 & 0.2509& 0.7898 & 0.1702 \\ 
    w/o Mixed Query & 0.7342 & 0.2513& 0.7894 & 0.1707 \\
    \hline
    w/o Global Sequence & 0.7336 & 0.2512& 0.7889 & 0.1720 \\
    w/o Real-time Sequence & 0.7328 & 0.2521& 0.7883 & 0.1729 \\
    \hline
    \hline
    \end{tabular}
}
\end{table} 

\subsubsection{Tokenization Structure Analysis.} 
We compare five variants: (1) \textit{HeMix}: the full model; (2) \textit{w/o Fixed Query}: removes the learnable fixed queries, disabling context-invariant interest modeling while retaining the dynamic NS-token queries; (3) \textit{w/o Mixed Query}: replaces the entire Mixed Hetero Attention module with AutoSplit-style tokenization, ablating both fixed and dynamic queries; (4) \textit{w/o Global Sequence}: drops the global behavior input; (5) \textit{w/o Real-time Sequence}: drops the real-time behavior input. Table~\ref{exp:abl_token} reports the results, from which we draw three observations.

\textit{First, every component is necessary.} All four ablations degrade both tasks simultaneously on both AUC and Logloss, and the full HeMix attains the best value on every metric. The consistency across two heterogeneous objectives (CTR and CVR) indicates that the gains come from genuinely better user-interest representations rather than task-specific overfitting.

\textit{Second, mixed queries are complementary, not redundant.} Removing only the fixed queries costs $0.0005$ CTR-AUC and $0.0014$ Logloss, whereas ablating the whole Mixed Query module costs $0.0008$ CTR-AUC and $0.0018$ Logloss. Since the two query types are individually weaker than their combination, the context-aware queries (which adapt per request) and the context-invariant queries (which encode population-level behavioral regularities) capture non-overlapping evidence. This directly supports the motivation behind C1: context-aware attention alone leaves the intrinsic structure of user sequences under-exploited.

\textit{Third, sequence semantics matter more than sequence length.} Dropping the real-time sequence causes the single largest degradation among all tokenization ablations ($-0.0022$ CTR-AUC / $-0.0019$ CVR-AUC), exceeding even the removal of the global sequence ($-0.0014$ / $-0.0013$). This is notable because the global sequence is over $30\times$ longer on average ($332$ vs.\ $10$ behaviors, in Table~\ref{tab:dataset}): a short but temporally fresh sequence carries more predictive signal than a long historical one. In LBS scenarios, a user's intent is tightly coupled to the current spatio-temporal context and shifts within a single session, so real-time behaviors are indispensable. This asymmetry empirically justifies our design choice of modeling the two sequences with \textit{separate} query groups and attention pipelines rather than concatenating them into one homogeneous sequence.

\begin{table}
\centering
\caption{Ablation analysis of interaction structure. }
\label{exp:abl}
\setlength{\tabcolsep}{2.0mm}
\scalebox{0.80}{
    \begin{tabular}{cc|cccc} 
    \hline
    \hline
    \multirow{2}{*}{}         & \multirow{2}{*}{Variants} & \multicolumn{2}{c}{CTR} & \multicolumn{2}{c}{CVR}  \\ 
    \cline{3-6}
    && AUC    & Logloss        & AUC    & Logloss         \\ 
    \hline
    \multirow{3}{*}{\begin{tabular}[c]{@{}c@{}}Interaction\\Structure\end{tabular}} 
    & HeteroMixer(Ours)      & \textbf{0.7350} & \textbf{0.2495}         & \textbf{0.7902} & \textbf{0.1696}          \\
    & RankMixer~\cite{zhu2025rankmixer} & 0.7345 & 0.2507         & 0.7898 & 0.1699          \\
    & Transformer~\cite{vaswani2017attention}   & 0.7336 & 0.2524         & 0.7884 & 0.1720          \\ 
    \hline
    \hline
    \end{tabular}
}
\end{table}

\subsubsection{Interaction Structure Analysis.} 
Keeping the tokenization layer fixed, we replace only the interaction block with three alternatives: (1) \textit{HeteroMixer Block} (ours), with heterogeneous token mixing and per-token HeteroFFN; (2) \textit{RankMixer Block}~\cite{zhu2025rankmixer}, with parameter-free multi-head token mixing and per-token FFN; and (3) \textit{Transformer Block}~\cite{vaswani2017attention}, with multi-head self-attention and a shared FFN. Table~\ref{exp:abl} reveals a clear two-level hierarchy.

\textit{First}, both mixing-based blocks (i.e., HeteroMixer and RankMixer) clearly outperform the Transformer block ($+0.0014$ and $+0.0009$ CTR-AUC, respectively). Since all three variants receive exactly the same tokens, this gap is attributable purely to the interaction operator: homogeneous pairwise attention, which computes a single shared similarity function over tokens drawn from semantically disjoint domains (user profiles, item attributes, contextual signals, behavioral interests), is a poor inductive bias for industrial ranking features. Crucially, both mixing-based blocks obtain this advantage without computing any explicit token-to-token similarity, showing that pairwise attention is not a prerequisite for high-quality interaction over heterogeneous tokens.

\textit{Second}, HeteroMixer further surpasses RankMixer by $0.0005$ CTR-AUC and $0.0012$ Logloss, with a consistent margin on CVR. The two blocks share the same fuse-and-reconstruct skeleton and differ only in the middle stage: RankMixer performs a parameter-free split-and-shuffle, which is a purely \textit{linear} rearrangement, whereas HeteroMixing applies \textit{head-specific low-rank MLPs} to each mixed token. The improvement therefore isolates the value of injecting non-linear, heterogeneity-aware transformation into the mixing stage, confirming that different semantic sub-spaces benefit from learning their own interaction patterns, precisely the gap identified in C2.

\subsubsection{Comparative Analysis.} Cross-referencing Tables~\ref{exp:abl_token} and~\ref{exp:abl} allows us to quantify the relative weight of the two pillars under an identical backbone. Removing Mixed Query (C1) costs $0.0008$ CTR-AUC, while replacing HeteroMixer with a Transformer (C2) costs $0.0014$, roughly $1.75\times$ the impact. Two conclusions follow. First, \textit{the interaction operator is the dominant driver}: once tokens are available, how they interact matters more than how they were produced, which explains why simply lengthening or flattening behavior sequences (as in flattening-based tokenizers) yields limited returns. Second, \textit{the two designs are additive rather than redundant}: the full model dominates every single-component ablation on all four metrics, and neither pillar can recover the other's contribution.

\subsection{Scaling Study}
\label{sec:scaling_exp}
\begin{figure}
    \setlength{\abovecaptionskip}{0.2cm}
    \setlength{\belowcaptionskip}{-0.4cm}
\centering
    \subfloat[CTR\_AUC v.s. \#Param]{
        \begin{minipage}[t]{0.48\linewidth}
        \centering
        \includegraphics[width=0.99\textwidth]{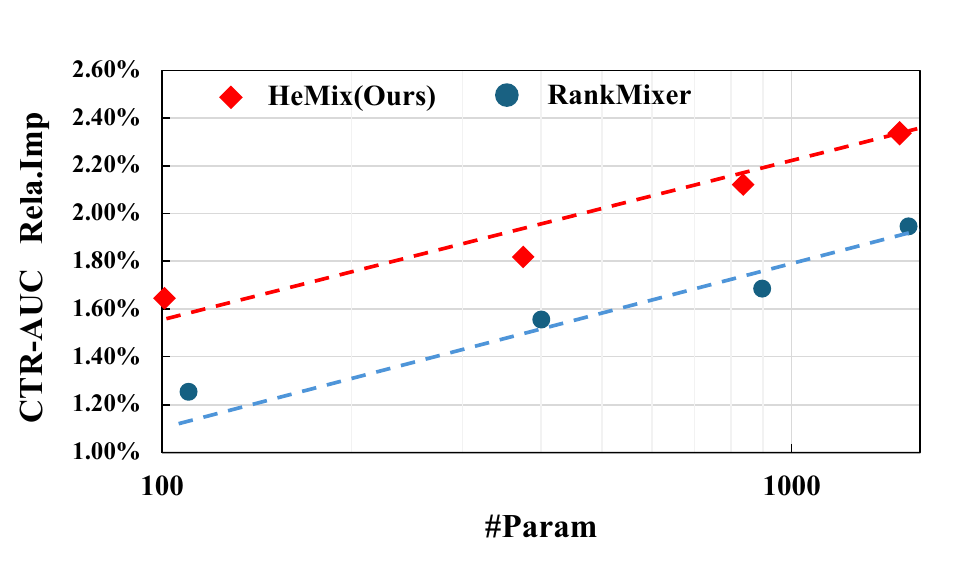}
        \end{minipage}
    }
    \centering
    \subfloat[CVR\_AUC v.s. \#Param]{
        \begin{minipage}[t]{0.48\linewidth}
        \centering
        \includegraphics[width=0.99\textwidth]{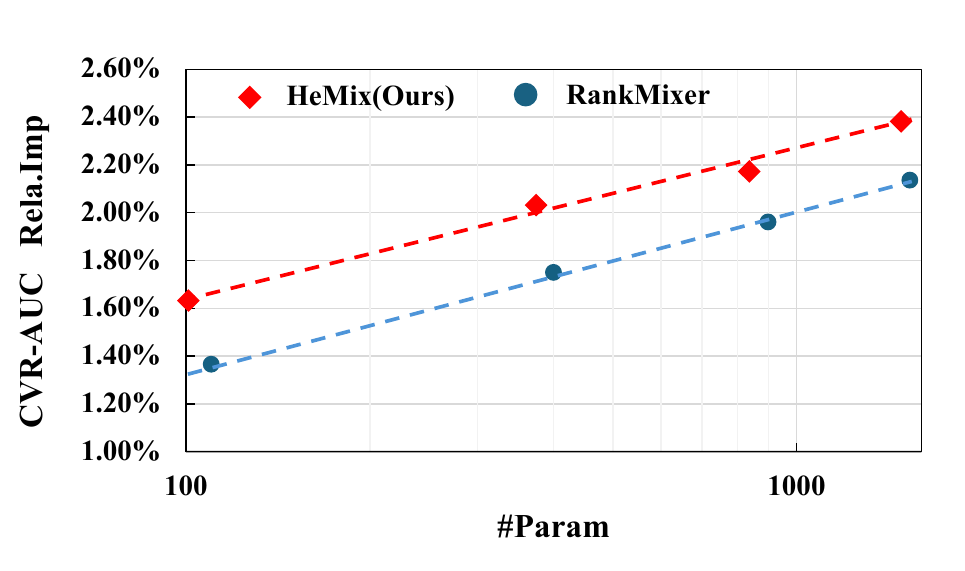}
        \end{minipage}
    }
    
    \subfloat[CTR\_AUC v.s. GFlops]{
        \begin{minipage}[t]{0.48\linewidth}
        \centering
        \includegraphics[width=0.99\textwidth]{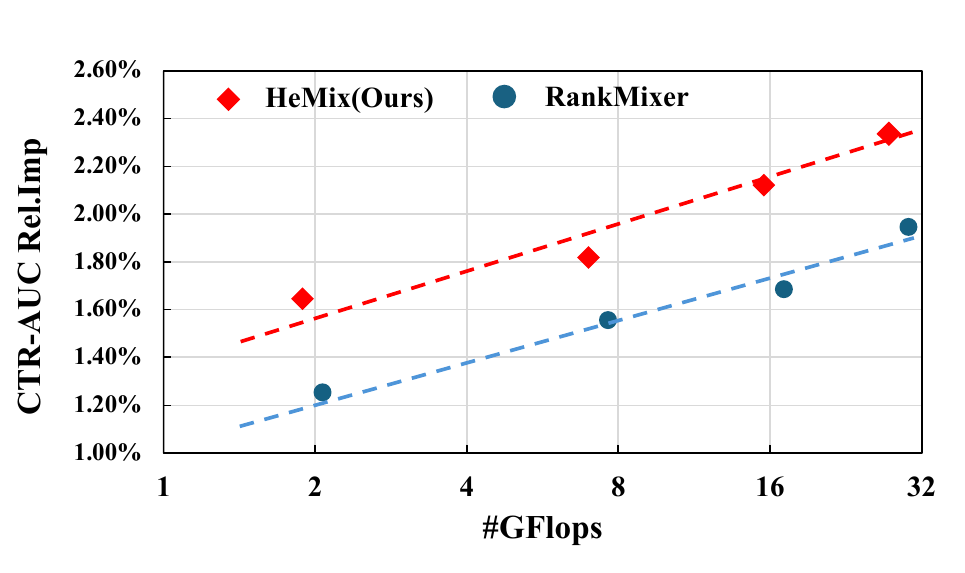}
        \end{minipage}
    }
    \centering
    \subfloat[CVR\_AUC v.s. GFlops]{
        \begin{minipage}[t]{0.48\linewidth}
        \centering
        \includegraphics[width=0.99\textwidth]{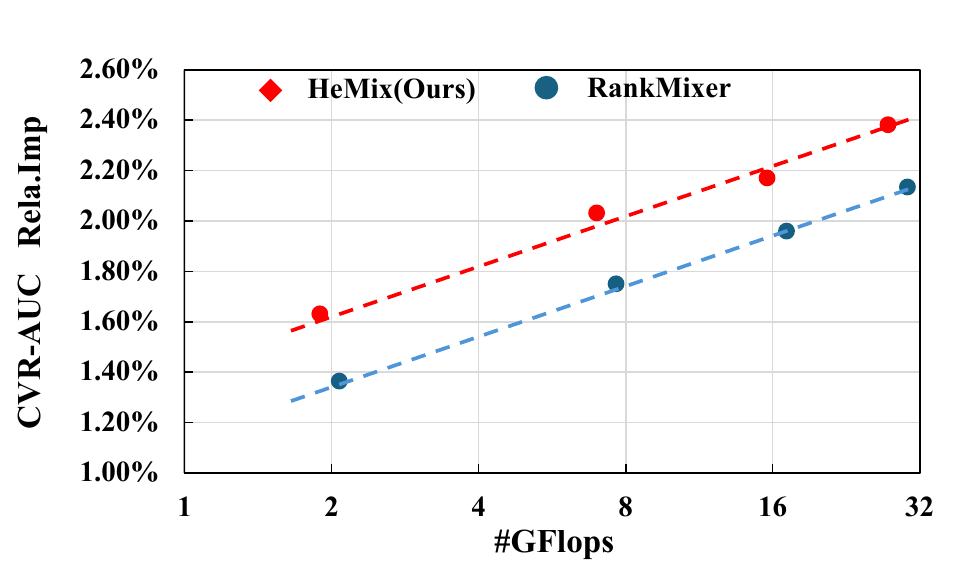}
        \end{minipage}
    }
\centering
\caption{Scaling behavior of HeMix vs.\ RankMixer.}
\label{fig:scaling}
\end{figure}

To evaluate the scalability of HeMix, we progressively scale the model from ${\sim}100$M to ${\sim}1500$M parameters by jointly expanding the block depth $L$ and the token dimension $d_T$, and compare against RankMixer under matched \#Params and GFLOPs budgets. Figure~\ref{fig:scaling} plots the relative CTR/CVR-AUC improvement over DLRM against both \#Params (a--b) and GFLOPs (c--d) on a logarithmic $x$-axis. Four findings emerge.

\textit{(i) Both architectures follow a clean log-linear scaling law.} For both models and both tasks, the measured points align closely with a straight line on the log-$x$ axis, i.e., $\textit{Rela.Imp} \approx \alpha \log_{10} {(\#
Params}) + \beta$. This confirms that the token-mixing paradigm inherits the predictable power-law behavior observed in language and vision models, and---more importantly for practice---that offline gains at a target budget can be \textit{extrapolated} rather than discovered by trial and error. The same log-linearity holds when the $x$-axis is compute (Figures~\ref{fig:scaling}(c) and Figures~\ref{fig:scaling}(d)), indicating the trend is not an artifact of a particular parameter--FLOP conversion.

\textit{(ii) HeMix dominates RankMixer through a persistent offset rather than a transient advantage.} HeMix lies strictly above RankMixer at every operating point, and the two fitted lines are nearly \textit{parallel}: over the ${\sim}15\times$ scaling range, both models gain about $0.70$ percentage points of relative CTR-AUC, while HeMix maintains a stable lead of ${\sim}0.4$ pp on CTR and ${\sim}0.25$ pp on CVR. This parallelism is a stronger result than a one-off win at a single scale. It rules out the two most common failure modes of architectural innovations in ranking: the advantage is neither a small-model artifact that vanishes once capacity becomes abundant, nor a large-model-only effect that is irrelevant to deployable budgets. Instead, the heterogeneity introduced by Query-Mixed Interest Extraction (C1) and HeteroMixing (C2) supplies information that additional parameters alone cannot recover, and therefore remains valuable at every scale.

\textit{(iii) The offset translates into an order-of-magnitude efficiency multiplier.} Because the two curves are parallel, a vertical gap is equivalent to a large horizontal gap. Concretely, HeMix at ${\sim}101$M parameters and $1.89$ GFLOPs already matches the CTR-AUC that RankMixer only reaches at ${\sim}850$M parameters and ${\sim}16$ GFLOPs, an ${\sim}8\times$ saving in parameters and compute for equal accuracy. Equivalently, at a fixed budget HeMix converts the same resources into strictly higher quality. For an industrial system where serving cost scales with FLOPs, this is the practically decisive form of the comparison: the same quality is delivered at roughly one order of magnitude lower cost.

\textit{(iv) Compute grows proportionally with parameters, empirically validating the design analysis.} Scaling HeMix from ${\sim}101$M to ${\sim}1484$M increases parameters by $14.7\times$ and GFLOPs by $14.6\times$, i.e., cost grows \textit{linearly} rather than quadratically in capacity. This directly corroborates the analysis in Section~\ref{sec:scaling}: the parameter count is dominated by HeteroFFN ($\#\text{Params}\approx 2kLNd_T^2$), while HeteroMixing contributes only $2LNd_Td_r$ under low-rank decomposition and never introduces an $\mathcal{O}(N^2)$ term. Consequently, HeMix can be scaled along $L$ and $d_T$ without triggering the attention-induced cost explosion that constrains Transformer-style rankers.

\begin{table}[t]
    \centering
    \caption{Relative improvements of HeMix over DLRM and RankMixer in online A/B testing on the AMAP platform.} 
    \label{exp:online_effi} 
    \setlength{\tabcolsep}{3.0mm}
    \scalebox{0.99}{
    \begin{tabular}{c|ccc}
    \hline
    \hline
    Base & GMV & PV\_CTR & UV\_CVR  \\
    \hline
    HeMix v.s. DLRM &3.61\% & 2.78\% & 2.12\% \\
    HeMix v.s. RankMixer  &0.88\%	&2.74\%	& 0.84\% \\ 
    \hline
    \hline
    \end{tabular}
    }
\end{table}
\subsection{Online Performance}

To evaluate the real-world impact of HeMix, we deploy the 100M-parameter variant in a large-scale online A/B test on AMAP's location-based service (LBS) recommender system, one of the largest production recommendation platforms, serving hundreds of millions of daily active users. We compare HeMix against two representative baselines: (1) the widely deployed DLRM production model, which has been serving the majority of AMAP traffic; and (2) RankMixer, a recent scaling-oriented architecture that has already demonstrated strong offline and online performance. We report the relative improvements on three core business metrics: 1) GMV (Gross Merchandise Value), reflecting overall transaction volume and platform revenue; 2) PV\_CTR (Page View Click-Through Rate), measuring user engagement at the impression level; and 3) UV\_CVR (Unique Visitor Conversion Rate), indicating the effectiveness of driving actionable outcomes per unique user.

Table~\ref{exp:online_effi} summarizes the online A/B test results. HeMix delivers consistent and statistically significant gains across all three metrics. Compared with the production DLRM, HeMix achieves substantial improvements of \textbf{+3.61\% GMV, +2.78\% PV\_CTR, and +2.12\% UV\_CVR}. These gains are particularly meaningful because they are obtained under strict online latency and serving-cost constraints, demonstrating that HeMix's efficiency-oriented design does not sacrifice business value. More importantly, even compared with the recent strong baseline RankMixer, HeMix still yields non-trivial improvements of \textbf{+0.88\% GMV, +2.74\% PV\_CTR, and +0.84\% UV\_CVR}. This confirms that the heterogeneous sequence tokenization and interaction mechanisms in HeMix provide additional online value beyond prior scaling-oriented architectures.

\section{Conclusion}

In this paper, we present \textbf{HeMix}, a scalable ranking model for industrial recommender systems that addresses two bottlenecks of existing ranking architectures: sequence tokenization that fails to jointly capture context-aware and context-invariant user interests from heterogeneous behavior sources, and feature interaction that is both computationally expensive and semantically homogeneous. HeMix introduces a Query-Mixed Interest Extraction module, which leverages dynamic context-aware queries together with fixed learnable queries to distill compact, semantically rich tokens from the global and real-time behavior sequences. On top of these tokens, stacked HeteroMixer blocks model high-order dependencies at a cost linear in the number of tokens, by pairing HeteroMixing for low-complexity heterogeneous cross-token interaction with HeteroFFN for token-specific non-linear transformation. Offline experiments at both the ${\sim}100$M and ${\sim}1500$M scales show consistent gains over state-of-the-art baselines at lower computational cost, and ablations confirm that the two designs are complementary. Fully deployed on the AMAP APP, HeMix delivers +0.88\% GMV, +2.74\% PV\_CTR and +0.84\% UV\_CVR online.

\bibliographystyle{ACM-Reference-Format}
\bibliography{hemix_bib}

\end{document}